\documentclass[11pt,letterpaper]{softmax_preprint}

\usepackage{tabularx}
\usepackage{textcomp}
\usepackage{pgfplots}
\usepackage{pgfplotstable}
\pgfplotsset{compat=1.18}
\definecolor{rqlevel1}{HTML}{F98981}
\definecolor{rqlevel2}{HTML}{D86B65}
\definecolor{rqlevel3}{HTML}{B74E4A}
\definecolor{rqlevel4}{HTML}{96302E}
\definecolor{rqlevel5}{HTML}{6A000B}
\newcommand{\rqhighx}{112}
\newcommand{\rqnx}{126}
\tikzset{
  rq note/.style={anchor=center,font=\scriptsize,text=black!80,inner sep=1pt},
  rq head/.style={anchor=south,align=center,font=\scriptsize\bfseries,text=black!80,inner sep=1.5pt},
}
\pgfplotsset{rq stacked/.style={
  xbar stacked, bar width=5.5pt,
  width=.5\linewidth, scale only axis,
  xmin=0, xmax=100, enlargelimits=false,
  y dir=reverse,
  axis x line*=bottom, axis y line*=left,
  axis line style={black!35},
  y axis line style={draw=none},
  xtick={0,25,50,75,100},
  xticklabel={\pgfmathprintnumber{\tick}\%},
  xmajorgrids, grid style={black!8,line width=.4pt},
  tick align=outside, major tick length=2pt,
  tick label style={font=\scriptsize,text=black!65},
  yticklabel style={anchor=east,font=\scriptsize,text=black!85},
  ytick style={draw=none},
  legend columns=5,
  legend style={at={(0.5,0)},anchor=north,yshift=-13pt,draw=none,fill=none,
    font=\scriptsize,text=black!80,column sep=6pt,
    /tikz/every even column/.append style={column sep=8pt}},
  legend image code/.code={\fill[#1] (0cm,-0.09cm) rectangle (0.3cm,0.09cm);},
  legend cell align=left,
  clip=false,
  every axis plot/.append style={draw=white,line width=.8pt},
}}
\usetikzlibrary{positioning,arrows.meta,calc,fit,backgrounds}

\setcitestyle{numbers,square,comma}
\colorlet{rqboxaccent}{WolfpackRed}
\colorlet{rqboxbg}{abstractbg}

\newtcolorbox{rqquestionbox}{%
  enhanced, breakable,
  colback=rqboxbg, colframe=rqboxbg,
  boxrule=0pt, arc=3pt, outer arc=3pt,
  left=8pt, right=8pt, top=1pt, bottom=1pt,
  fontupper=\small, before skip=8pt, after skip=4pt
}
\newcommand{\rqquestion}[2]{%
  \begin{rqquestionbox}
    \textbf{\textcolor{rqboxaccent}{\faQuestionCircle[regular]\,RQ#1: #2}}
  \end{rqquestionbox}%
}

\newtcolorbox{rqsummarybox}{%
  enhanced, breakable,
  colback=rqboxbg, colframe=rqboxbg,
  boxrule=0pt, arc=3pt, outer arc=3pt,
  boxsep=0pt,
  left=8pt, right=8pt, top=3pt, bottom=3pt,
  fontupper=\small, before skip=10pt, after skip=10pt
}
\newcommand{\rqsummary}[2]{%
  \begin{rqsummarybox}
    \textbf{\textcolor{rqboxaccent}{\faClipboardCheck\,RQ#1 summary.}}~#2
  \end{rqsummarybox}%
}

\renewcommand{\finding}[2]{%
  \par\addvspace{\medskipamount}%
  \Needspace{5\baselineskip}%
  \noindent\begin{minipage}{\linewidth}%
    \setlength{\parskip}{0pt}%
    {\bfseries\textcolor{rqboxaccent}{\faLightbulb[regular]\,Finding~#1.}~#2\par}%
  \end{minipage}\par\nobreak
  \smallskip\noindent\ignorespaces
}

\newtcolorbox{participantquotebox}{%
  enhanced, frame hidden, sharp corners,
  colback=rqboxbg!45!white,
  fontupper=\small,
  borderline west={1.5pt}{0pt}{rqboxaccent!70},
  boxsep=0pt, left=22pt, right=9pt, top=3pt, bottom=3pt,
  before skip=7pt, after skip=7pt,
  overlay={\node[anchor=north west, inner sep=0pt, text=rqboxaccent!70]
    at ([xshift=6pt,yshift=-1pt]interior.north west) {\small\faQuoteLeft};}
}
\newcommand{\participantquote}[2]{%
  \begin{participantquotebox}
    \textit{#2}\unskip\nobreak\hfill\mbox{\quad\scriptsize\textcolor{rqboxaccent}{#1}}\par
  \end{participantquotebox}%
}

\newlength{\panelsep}
\newlength{\panelgap}
\newlength{\panelrun}
\newlength{\panelstep}
\newlength{\marker}
\newlength{\itemlead}

\title{From Delegation to Action Approval: \\ An Exploratory Study of Developers' Practices in Collaborating with AI Coding Agents}
\runningtitle{From Delegation to Action Approval}

\author{Yinan Wu\affmark{1}, Dominik Wermke\affmark{1}, Ze Shi Li\affmark{2},
  Kathryn Thomasset Stolee\affmark{1}, Bowen Xu\affmark{1}}
\affiliation{%
  \affmark{1}North Carolina State University, Raleigh, NC, USA\\
  \affmark{2}University of Oklahoma, Norman, OK, USA}
\authornote{}

\emails{\email{ywu92@ncsu.edu},\quad\email{dwermke@ncsu.edu},\quad\email{zeshili@ou.edu},\quad\email{ktstolee@ncsu.edu},\quad\email{bxu22@ncsu.edu}}

\logo{\includegraphics[height=1.0cm]{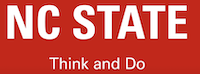}}
\runninglogo{\includegraphics[height=0.65cm]{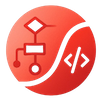}}
\banner{%
  \includegraphics[height=0.85cm]{assets/logo.png}%
  \hspace{0.25cm}%
  \raisebox{0.25cm}{Soft\textcolor{bannerlight}{Max} Lab}}

\begin{document}
\maketitle

\begin{abstract}
Developers now collaborate with AI coding agents that write code and take actions on their behalf. Beyond reviewing this code, developers decide what to delegate and which agent actions to approve. Some of these actions, such as running shell commands or reading files, can have effects that never appear in the resulting code. How developers make these decisions and how much effort this requires are not well understood. To better understand this collaboration, we surveyed 203 developers who use AI coding agents. We found that careful code review does not necessarily extend to action approval: 61\% of developers who carefully reviewed most or all generated code still often approved at least one type of agent action without close review. Developers aware of security or privacy incidents or concerning practices did so more often than those who were not. We also found that autonomy varied more between developers than between development activities. Although developers reported a lighter workload, supervising agents remained mentally demanding. When asked what would make them grant more autonomy, developers wanted more than reliability: more than half of their responses called for agents that are bounded, observable, or reversible. Our findings suggest that as agents take on more of the coding, the challenge shifts toward supporting the developers who supervise them.
\end{abstract}

\section{Introduction}
\label{sec:intro}

Many developers now hand software tasks to artificial intelligence (AI) coding agents.\footnote{In this paper, we use ``AI coding agents,'' ``coding agents,'' and ``agents'' interchangeably to refer to AI tools that use large language models (LLMs) and can call external tools to carry out software development tasks.}
In the 2025 Stack Overflow Developer Survey, 31\% of developers reported using AI agents~\cite{stackoverflow2025survey}, and a smaller follow-up survey in April 2026 put the figure at 59\%~\cite{stackoverflow2026pulse}.
Consider a typical moment in this work:
midway through a task, an agent stops and asks for permission to run a shell command or modify the codebase, and the developer approves or rejects the action.
Agents now act on developers' behalf, yet developers remain collaborators rather than bystanders.
They decide what to delegate, how independently agents may work, and when to intervene, as well as which actions to approve and whether to accept the resulting code.

Prior studies show that developers stay involved after handing work to agents.
Experienced developers, for example, plan with agents and supervise them closely to protect software quality~\cite{huang2025professional}.
Much of this care, however, focuses on the code an agent produces.
Coding agents introduce another target for assurance: the actions they take while producing that code.
A command can delete data and a file read can expose secrets, even when these effects leave no trace in the resulting code.
Prior work on permission prompts shows that users may approve requests without much attention~\cite{felt2012android}, while records from millions of Claude Code sessions show that experienced users allow automatic approval more often while also interrupting the agent more often~\cite{anthropic2026autonomy}.
As agents take on more work, human involvement therefore does not disappear; it shifts toward deciding what to delegate, when to intervene, and what to check.

These decisions are closely related but have often been studied separately.
Research on human-AI collaboration suggests that effective collaboration requires people to direct the AI, predict what it will do, and keep the effort of working with it manageable~\cite{klien2004ten}.
For coding agents, these needs raise three questions.
First, how much autonomy do developers give their agents, and when do they intervene?
Second, does the care developers give to generated code also extend to the actions agents ask to perform?
Third, what does supervising agents demand from developers, and what would help them delegate more?
Together, these questions concern how developers \emph{orchestrate}, \emph{assure}, and \emph{experience} their work with coding agents.

To answer these questions, we surveyed 203 developers who use AI coding agents, recruited through a community channel and Prolific.
Our findings reveal three tensions in developer-agent collaboration.
First, \textbf{autonomy is developer-dependent, and intervention centers on keeping agents aligned with intent}.
Reported autonomy varied much more across developers than across development activities, while developers continued to intervene when agents misunderstood the task, chose an unsuitable approach, or changed more than requested.
Second, \textbf{careful code review does not necessarily extend to careful action approval}.
Among respondents who carefully reviewed most or all generated code, 61\% often approved at least one type of agent action without close review.
Higher reported autonomy was also associated with more frequent approval without close review, highlighting code and agent actions as distinct assurance targets.
Third, \textbf{workload savings coexist with demanding supervision}.
Although most developers reported that agents reduced their workload, many still found supervising them mentally demanding.
Most developers wanted agents to complete their work reliably, and to give agents more autonomy, more than half also wanted autonomy that is bounded, observable, and reversible.

Our work makes three contributions:
\begin{itemize}
  \item We provide an empirical characterization of how developers orchestrate, assure, and experience work with AI coding agents, based on a survey of 203 active coding-agent users.
  
  \item We distinguish code review from action approval as two forms of assurance, showing that careful code review does not necessarily extend to careful approval of agent actions.
  
  \item We characterize the supervision demands developers experience and identify what developers need before giving agents more autonomy: agents that are reliable and also bounded, observable, and reversible.
\end{itemize}

\section{Related Work}
Research on human-AI collaboration treats AI agents as partners in joint work rather than simply as tools~\cite{hauptman2023adapt,klien2004ten}. To work well with such a partner, people need to direct it, predict what it will do, and keep the effort of working with it low~\cite{klien2004ten}. Coding agents bring these needs into software development by carrying out multi-step work on developers' behalf. We review related work in three areas that match these needs: how developers delegate work to agents and supervise it, how they check the code and actions of agents, and how they experience working with agents.

\subsection{Delegation and Supervision of Coding Agents}

Delegating work to a coding agent requires deciding what it may do without asking. Automation research describes levels of human control, from selecting suggested actions to allowing independent execution~\cite{parasuraman2000model}. Feng et al.~\cite{feng2025levels} applied this idea to agents through five levels defined by the user's role: operator, collaborator, consultant, approver, and observer. Choudhuri et al.~\cite{choudhuri2026you} surveyed 448 Microsoft developers about the levels of AI autonomy they would accept in software engineering work. Accepted autonomy varied across tasks: it was lower when developers were responsible for the outcome and higher for demanding tasks. It also varied across developers and was higher among those with more AI experience and tolerance for risk.
After delegating, developers keep guiding the agent and checking its work. Studies of coding agents describe this supervision as planning with the agent, setting limits, monitoring progress, and reviewing results~\cite{huang2025professional,dhanorkar2026human,park2026work}. During supervision, developers may step in when they notice a problem, or the agent may stop and ask. Deciding when an agent should ask involves a tradeoff: asking interrupts the developer, but continuing on a wrong assumption can create work to fix later~\cite{horvitz1999principles}.
Two questions remain open. First, Choudhuri et al.~\cite{choudhuri2026you} asked which levels of autonomy developers would accept, not how much autonomy they grant to the agents they already use, or whether this depends more on the developer or on the activity. Second, prior work explains why an agent should sometimes stop and ask, but says less about where developers want it to stop. Our survey addresses both gaps: we measured how much autonomy developers report granting across development activities, and we compared why developers intervene with when they want agents to check in.

\subsection{Code Review and Action Approval}

Coding agents create two targets for assurance: the code they produce and the actions they take while producing it. Code review is a standard practice for assuring software quality~\cite{sadowski2018modern}, but Bacchelli and Bird~\cite{bacchelli2013expectations} found that reviewers struggled most with understanding the changes they reviewed. Dhanorkar et al.~\cite{dhanorkar2026human} reported the same difficulty with agents: developers found unfamiliar agent-generated code hard to review and sometimes relied on passing tests as evidence of correctness. Developers may also trust AI-generated code too much: participants with an AI assistant wrote less secure code, yet were more likely to believe it was secure~\cite{perry2023users}.
Agent actions create a different problem, because their effects may never appear in the resulting code. During execution, an agent may ask permission to run commands, access files, install dependencies, or use other resources. An agent that reads untrusted content can also be steered by instructions hidden in that content~\cite{greshake2023not}. Research on permission prompts shows that displaying a request does not ensure that users understand it~\cite{felt2012android}, and that repeated warnings receive less attention over time~\cite{anderson2015polymorphic}. More generally, people monitor automation less closely when busy with other tasks~\cite{parasuraman2010complacency}. Usage data from coding agents show that approval habits also change with experience: experienced Claude Code users enabled automatic approval more often, but also interrupted agents more often~\cite{anthropic2026autonomy}.
These two lines of work have developed separately: code review studies focus on developers checking code changes, and permission studies focus on users responding to prompts, mostly outside software development. Coding agents put both decisions in front of the same developer within one task. We know little about whether developers who review an agent's code carefully also look closely at the actions it asks to perform, or whether developers who grant more autonomy or know more about security risks approve actions more carefully. Our survey asked each respondent about both practices, so we can compare them within the same person and relate them to developers' other practices and experiences.

\subsection{Developer Experience with AI Coding Assistants}

Developers often feel that AI coding assistants help them more than measurements show. Most participants in one study preferred Copilot even though it did not improve completion time or task success~\cite{vaithilingam2022expectation}, and experienced open-source developers believed AI tools sped them up while they took 19\% longer to complete tasks~\cite{becker2025measuring}. Part of the cost is verification: programmers using Copilot spent about a fifth of their time verifying its suggestions~\cite{mozannar2024reading}, a case of automation turning work into demanding monitoring~\cite{bainbridge1983ironies}. These costs are not only about time: Wu et al.~\cite{wu2026developers} reported guilt and self-doubt among some developers who relied on AI. Frequent users of AI programming tools trusted them more, but also felt a greater threat to their job security~\cite{kuhail2024will}. Knowledge workers who were more confident in generative AI reported less critical thinking when using it~\cite{lee2025impact}.
Much of this evidence comes from code completion tools such as Copilot. Coding agents widen what developers must supervise: instead of checking single suggestions, developers may need to understand changes across files and sessions, monitor multi-step work, decide when to intervene, and respond to approval requests. We treat the resulting challenges as part of developers' experience with agents. Less is known about which supervision activities developers find demanding and how these demands relate to their views of their work. We measured how demanding developers found each supervision activity and compared these ratings with their views of their work.
These challenges also point to how agents should change. Choudhuri et al.~\cite{choudhuri2026copilot} surveyed 860 Microsoft developers and identified 22 AI systems they wanted built. This study asked what developers want built, but not what would make them willing to give agents more autonomy. We asked both: what one thing developers would fix or add to their agents, and what would make them willing to give agents more autonomy.

\section{Methodology}
We conducted an exploratory survey study to understand how developers
orchestrate, assure, and experience their work with AI coding agents.
\begin{figure}[tb]
\centering
\setlength{\panelrun}{\dimexpr\linewidth-10\panelsep-4\panelgap-2pt\relax}
\setlength{\panelstep}{\dimexpr2\panelsep+\panelgap\relax}
\begin{tikzpicture}[
  every node/.append style={execute at begin node=\hyphenpenalty 10000\exhyphenpenalty 10000\relax},
  content/.style={align=left, anchor=north west, inner sep=0pt},
  panel/.style={fill=rqboxbg, rounded corners=3pt, inner sep=\panelsep},
  arr/.style={-{Stealth[length=1.6mm]}, line width=.7pt, draw=rqboxaccent},
]
\newcommand{\headfont}{\fontsize{8.2}{9.9}\selectfont}
\newcommand{\itemfont}{\fontsize{7.2}{8.5}\selectfont}
\setlength{\marker}{6.5pt}       
\setlength{\itemlead}{4pt}       
\newcommand{\phead}[2]{%
  {\headfont\bfseries\color{rqboxaccent}#1\hspace{4pt}#2\par}%
  \nointerlineskip\vspace{4pt}%
  {\color{rqboxaccent!30}\rule{\linewidth}{0.4pt}}\par}
\newcommand{\pitem}[2]{%
  \vspace{\itemlead}{\itemfont\noindent
    \makebox[\marker][l]{\textbullet}#1\\\hspace*{\marker}{\color{black!55}#2}\par}}

\node[content, text width=0.2187\panelrun] (c1) at (0,0) {%
  \phead{\faQuestionCircle[regular]}{Research Questions}
  \pitem{{\bfseries\color{rqboxaccent}RQ1}~Orchestration}{Delegation and autonomy}
  \pitem{{\bfseries\color{rqboxaccent}RQ2}~Assurance}{Code review and approval}
  \pitem{{\bfseries\color{rqboxaccent}RQ3}~Experience}{Workload and supervision}};
\node[content, text width=0.1941\panelrun] (c2) at ([xshift=\panelstep]c1.north east) {%
  \phead{\faClipboardList}{Survey Design}
  \pitem{Question development}{Literature, practitioners}
  \pitem{Pilot testing}{3 think-aloud, 15 pilot}
  \pitem{Final questionnaire}{7 sections, 20--30 min}};
\node[content, text width=0.2072\panelrun] (c3) at ([xshift=\panelstep]c2.north east) {%
  \phead{\faUsers}{Recruitment}
  \pitem{Non-probability sampling}{Community, Prolific}
  \pitem{Eligibility}{Adults using agents}
  \pitem{Five weeks}{US\$6 compensation}};
\node[content, text width=0.1757\panelrun] (c4) at ([xshift=\panelstep]c3.north east) {%
  \phead{\faFilter}{Screening}
  \pitem{Rule-based screening}{583 $\rightarrow$ 210}
  \pitem{Flagging and\\\hspace*{\marker}manual review}{7 of 38 excluded}
  \vspace{\itemlead}{\itemfont\noindent\hspace*{\marker}%
    \bfseries\color{rqboxaccent}203 responses\par}};
\node[content, text width=0.2043\panelrun] (c5) at ([xshift=\panelstep]c4.north east) {%
  \setlength{\itemlead}{6pt}
  \phead{\faChartBar[regular]}{Data Analysis}
  \pitem{Quantitative analysis of\\\hspace*{\marker}closed-ended questions}{Descriptives, associations}
  \pitem{Qualitative analysis of\\\hspace*{\marker}open-ended questions}{STGT, three questions}};

\path (c1.north) coordinate (top);
\path let \p1=(c1.south), \p2=(c2.south), \p3=(c3.south), \p4=(c4.south), \p5=(c5.south)
  in coordinate (bot) at (0,{min(\y1,\y2,\y3,\y4,\y5)});
\begin{scope}[on background layer]
  \foreach \i in {1,...,5} {
    \node[panel, fit=(c\i.west |- top) (c\i.east |- bot)] (p\i) {};
  }
\end{scope}

\foreach \i [evaluate=\i as \j using int(\i+1)] in {1,...,4} {
  \draw[arr] (p\i.east) -- (p\j.west);
}
\end{tikzpicture}
\caption{Study overview. The research questions guide survey design, recruitment, response screening, and data analysis.}
\label{fig:method-overview}
\end{figure}
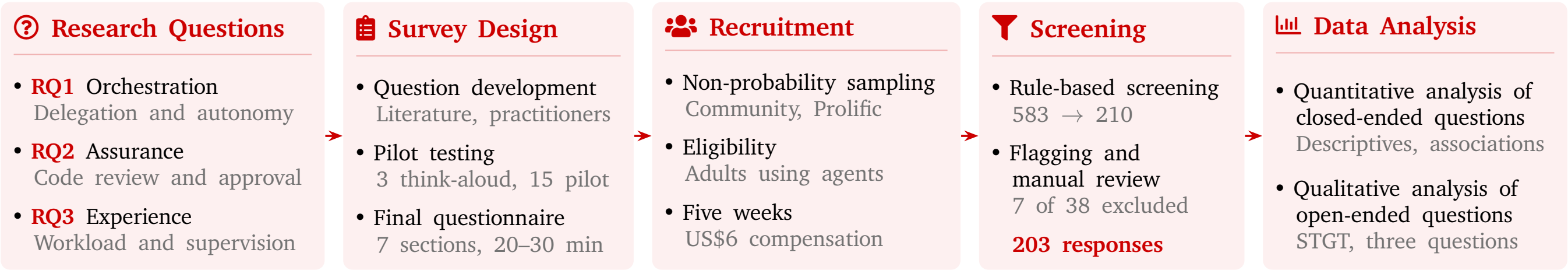
Figure~\ref{fig:method-overview} summarizes the study process.
Software engineering survey guidelines~\cite{linaaker2015guidelines}
informed the survey design. We organized the survey and analyses around
three research questions and proceeded through questionnaire development
and piloting, participant recruitment, response screening,
and quantitative and qualitative analysis.

\subsection{Research Questions}
\label{sec:rq_overview}

\rqquestion{1}{How do developers orchestrate their work with AI coding agents?}
\noindent

\noindent Agents can carry out multi-step tasks on their own, so developers decide how much to hand over and when to step in. Prior work asked which levels of autonomy developers would accept~\cite{choudhuri2026you}, not how much they grant in practice or whether that depends more on the developer or the activity. We estimate how much autonomy varies across developers and across activities and explore how developers delegate and intervene.

\rqquestion{2}{How do developers assure the work of AI coding agents?}
\noindent
Developers are responsible for both the code agents write and the actions they take. Code review checks the code, but actions such as running commands or reading secrets may never appear in a code change, and people often pay little attention to permission prompts~\cite{felt2012android}. We compare code review with action approval within the same respondents and relate approval to autonomy and risk awareness.

\rqquestion{3}{How do developers experience working with AI coding agents?}
\noindent
Agents may reduce developers' direct workload while introducing
new work in supervising and checking agent behavior. Checking AI output
already takes a substantial share of developers' time~\cite{mozannar2024reading},
and agents broaden what must be supervised. We examine the benefits
developers report alongside the mental demands of supervision and recurring
agent problems, and relate these demands and problems to developers' views
of their work. We then analyze what developers want improved and what
would make them willing to grant agents greater autonomy.

\subsection{Survey Design}
\label{sec:survey-design}

\noindent\textbf{Stage 1: Initial question development.}
We focused on developers' practices and experiences with coding agents.
To generate candidate topics for the questionnaire, we conducted a
formative scan of grey literature~\cite{garousi2019guidelines} in June
2026. We searched Google for ``developer AI coding agents'' combined with
each of three terms (\emph{use}, \emph{limitations}, and \emph{impact})
and reviewed the first 20 results per search. The purpose was to identify
topics developers discussed for questionnaire development, rather than to
conduct a systematic literature review. The first author coded each post
to identify an initial set of topics, and we consulted two industry
professionals about which topics they considered important. Their
feedback, further team discussions, and related literature~\cite{chen2026code, feng2025levels, mozannar2024reading, wu2026developers, huang2025professional, perry2023users, dhanorkar2026human} informed six
areas: delegation, interaction, verification, risk management, challenges,
and impact on developers. We developed questions for these areas and
refined them with feedback from a third industry professional. The
questions were study-specific measures for exploratory description rather
than validated psychological scales.

\noindent\textbf{Stage 2: Pilot testing and refinement.}
We first conducted think-aloud sessions with three participants to
identify unclear wording and difficulties understanding the survey.
Their feedback led us to revise wording, merge or remove questions, and
enable backward navigation. We also moved challenge-related questions
from a separate section into the relevant activity sections, allowing
participants to consider challenges in context and avoiding a long block
of negatively framed questions.

We then piloted the revised survey with another 15 participants and
collected feedback after completion. We used this feedback to further
revise the questions and response options. For example, the initial
autonomy question used a five-level framework based on users' roles when
interacting with agents~\cite{feng2025levels}. Participants found the
options difficult and time-consuming to interpret, so we replaced them
with a five-point autonomy scale ranging from close human supervision to
independent agent work. Participants also wanted to explain when and why
they would allow greater autonomy. We therefore added an open-ended
question in the delegation section asking what would make them willing to
give agents more autonomy. Pilot responses were excluded from the study
analyses.

\noindent\textbf{Stage 3: Final questionnaire.}
The final survey was implemented in Qualtrics~\cite{qualtrics} and
designed to take approximately 20--30 minutes. It contained seven
sections: general coding-agent use (\textsf{S1}), delegation
(\textsf{S2}), monitoring and intervention (\textsf{S3}), review and
verification (\textsf{S4}), risk management (\textsf{S5}), developer
experience (\textsf{S6}), and participant background (\textsf{S7}).
\textsf{S1} served as a warm-up before the main sections. We placed
background questions near the end because asking about personal
characteristics first may influence how participants answer later
questions~\cite{hughes2016rethinking}.

In addition to the autonomy question in \textsf{S2}, two optional
open-ended questions appeared at the end of the survey: one about
experiences not covered elsewhere and one about the improvement
participants most wanted from coding agents. The questionnaire included
two required attention checks and one required verification question
asking whether the respondent was completing the survey as a human or an
automated agent. These supported response-quality screening; all other
survey questions were optional.

\subsection{Participant Recruitment}
\label{sec:recruitment}

We used non-probability sampling~\cite{baltes2022sampling} to recruit
current coding-agent users over five weeks in August and September 2026.
Our goal was to capture a range of practices and experiences rather than
obtain a representative sample of developers. Participants had to be at
least 18 years old, currently use AI coding agents, and consent to the
study. We recruited through two channels. We first distributed the survey
through a community channel consisting of social media posts and
referrals, primarily via LinkedIn. Because this channel mainly reached
participants in North America, we then used Prolific to recruit
participants outside North America and broaden the sample's geographic
coverage. On Prolific, we used prescreening filters to restrict
participation to users whose employment role related to software
development and who reported using AI coding assistants. For both
channels, we confirmed current agent use with our own eligibility
question.
Participants with valid responses could receive US\$6 in compensation.
Prolific participants were paid through the platform; community-channel
participants could receive a US\$6 Amazon gift card. Community-channel
participants provided an email address only if they requested
compensation and could otherwise decline it.

\subsection{Response Screening}
\label{sec:screening}

We merged 583 submissions, 492 from the community channel and 91 from
Prolific, and applied two steps of quality screening. These steps
retained 210 and then 203 responses (124 from the community channel and
79 from Prolific).

\noindent\textbf{Step 1: Rule-based screening.}
Rule-based screening first excluded 122 of the 583 submissions: 38
incomplete submissions, 27 attention-check failures, 43 duplicates
(repeated email addresses or identical open-ended answers of at least
50 characters), 5 submissions that identified themselves as bots or
language models, and 9 for other reasons (pilot, test, ineligible,
withdrawn, or excessive missing answers). This left 461 responses.

Short completion times can indicate careless
responding~\cite{curran2016methods}. Following prior survey
studies~\cite{lambiase2026quantifying}, we used half the 24-minute pilot median (12 minutes) as an initial threshold. All Prolific responses
exceeded this threshold. Before applying a time threshold to the 379
remaining community-channel responses, however, we observed two groups
of completion times on visual inspection. Following work that models
rapid and effortful responding using response-time mixture
models~\cite{schnipke1997modeling,ulitzsch2022response}, we fitted a
two-component log-normal mixture to these times. The two-component model
fitted the data better than a single distribution (difference in Bayesian information criterion, $\Delta\mathrm{BIC}=121$), and
the two components were equally likely at 13.2 minutes. We therefore
excluded community-channel responses completed in under 13 minutes.
The time screening excluded 251 responses, all from the community
channel, leaving 210 responses.

\noindent\textbf{Step 2: Flagging and manual review.}
We examined the 210 retained responses for suspicious patterns requiring
further review. Three automated diagnostics flagged straightlining (the
same rating for every item in a rating matrix), high similarity in
structured answers across respondents, and near-identical open-ended
answers across respondents. Our flagging rules, including straightlining
four or more matrices, placed 33 cases in a manual-review queue.
Because AI agents can pass attention checks and produce human-like
answers~\cite{westwood2025potential, zhang2022beyond}, we used two additional LLM-based
flagging passes. Both used GPT-5.6 and reviewed all 210 anonymized
responses without access to identifiers. One followed our review protocol;
the other used a general rubric that the model drafted independently of
our protocol. Each returned recommendations with supporting evidence.
These outputs were used only to identify responses for human inspection;
no response was excluded based on an LLM judgment alone, since LLM
judgments can approach but not replace human
annotation~\cite{ahmed2025can}.
The protocol-based pass flagged 14 responses, all among the 33 cases
already flagged by the automated diagnostics. The independent pass
flagged nine responses, five of which were new. In total, 38 responses
underwent manual review. The research team examined the responses and
available evidence and excluded seven. For example, one excluded
response gave identical ratings across five matrices and included an
open-ended answer beginning ``\textit{A strong survey response would
be:}'', followed by a block quote. The final sample contained 203
responses.

\subsection{Data Analysis}
\label{sec:data-analysis}

We used quantitative analysis for closed-ended questions and qualitative
analysis for the three open-ended questions.

\noindent\textbf{Response handling.}
Each analysis used the available responses to its relevant questions or
items without imputing missing values. An unanswered multiple-choice
question was not treated as selecting no options. \emph{Not applicable}
and \emph{Not sure} responses were excluded from rating analyses. We
reviewed entries under \emph{Other} and merged them into an existing
option when their meaning clearly matched. Invalid entries were treated
as missing. The remaining \emph{Other} entries accounted for 0--3.0\%
of responses per question. We included them in descriptive summaries but
excluded them from association analyses because they represented
different answers.

\noindent\textbf{Quantitative analysis.}
Following survey-analysis guidance and
practices~\cite{kitchenham2008personal,kitchenham2003principles,liang2024large},
we summarized closed-ended responses using counts, percentages, and
distributions. Percentage denominators were the valid responses to each
question or item. Reported frequencies reflect respondents' perceptions
rather than directly observed behavior.
For presentation, we sometimes combined adjacent response categories,
such as \emph{Often} and \emph{Very often}, or \emph{Agree} and
\emph{Strongly agree}; statistical analyses retained the original
response scales. In the autonomy summaries, ratings of 4 or 5 on the
five-point scale were reported as high autonomy. We considered
supervision activities rated \emph{Very much} or \emph{Extremely} to
have high reported mental demand.
For association analyses, we selected statistical tests according to the
variables being compared. We used Spearman's
$\rho$~\cite{spearman1961proof} for two ordinal ratings; Mann--Whitney
$U$ tests~\cite{mann1947test} with rank-biserial correlation
$r_{\mathrm{rb}}$~\cite{kerby2014simple} for an ordinal rating between
two groups; Fisher's exact tests~\cite{fisher1922interpretation} with
$\phi$~\cite{cohen2013statistical} for two binary variables; and exact
McNemar tests~\cite{mcnemar1947note} for paired binary answers.
For comparisons of two groups on a binary outcome, we also report odds ratios with exact conditional 95\% confidence intervals. Confidence intervals for $\rho$ and $\phi$ are 95\% percentile intervals from 2,000 bootstrap resamples of respondents.
To control for multiple testing, we applied the Benjamini--Hochberg (BH)
correction~\cite{benjamini1995controlling} within each analysis set,
using adjusted $q<.05$ as the significance threshold. An analysis set
was a conceptually defined block of pairwise tests addressing the same
relationship, such as the $11\times7$ comparisons between development
activities and action types in Section~\ref{sec:rq2}. All
associations were treated as exploratory and were not interpreted as
causal relationships.

\noindent\textbf{Qualitative analysis.}
We used the basic data analysis procedures of Socio-Technical Grounded
Theory (STGT)~\cite{hoda2021socio,hoda2025qualitative} to analyze the
three open-ended questions. These procedures support developing emerging
categories about socio-technical phenomena, such as developers'
interactions with AI agents~\cite{wu2026developers}, without requiring
full theory development.
Coding was inductive: we did not start from predefined codes, although
we organized the resulting categories within the scope of our research
questions. The first author initially coded all responses. As coding
progressed, similar codes were compared and combined, their meanings were
refined, and related codes were grouped into broader concepts and
categories. The research team discussed the emerging categories, which
were further refined based on this feedback.

We excluded empty, non-substantive, and unrelated responses. A response
could receive multiple codes and contribute to multiple categories.
Category counts represent distinct included responses within each
question rather than the number of code assignments. We report the
categories that address RQ3 (Section~\ref{sec:rq3}), drawing on the two
questions about desired improvements and greater autonomy. The third
question asked about experiences not covered elsewhere and was used to
provide illustrative quotations.

\subsection{Ethical Considerations}
\label{sec:ethics}

The study was approved by the authors' institutional review board (IRB).
Participation was voluntary and required informed consent. Respondents
could withdraw their responses at the end of the survey. Contact emails
were collected only for compensation and stored separately from survey
responses. Only approved researchers could access the raw data. All
reported data were de-identified, and quotations are attributed to
participant codes.

\section{Results}
\label{sec:results}


\subsection{Respondent and Agent-Use Overview}
\label{sec:respondent-overview}

The final sample included 203 respondents: 124 from the community channel and 79 from Prolific. The 199 respondents who reported their country of residence lived in 25 countries, with most based in the United States (63.8\%). By continent, 129 lived in North America, 24 in Europe, 24 in Africa, 15 in Asia, five in South America, and two in Oceania.
Most respondents answered from the perspective of industry developers or interns (43.3\%) or freelance or independent developers (32.5\%), and 52.7\% reported working as software developers or engineers. Programming experience varied, with 3--5 years being the largest group (27.6\%). Respondents also worked in varied settings, including projects involving proprietary or confidential code (52.2\%), sensitive data (49.3\%), production systems (38.4\%), or safety-critical systems (32.0\%).
Respondents were active coding-agent users. They had used agents for a median of 12 months ($n=194$), and 50.5\% used them daily. The most frequently used agents were Claude Code (79.2\%), GitHub Copilot in agent mode (58.9\%), conversational AI used for coding (54.5\%), and OpenAI Codex (49.5\%). Most respondents used several agents: 91.1\% used at least two and 73.8\% at least three. They most often used agents to complete tasks faster (80.3\%), offload tedious or repetitive work (63.5\%), and explore unfamiliar codebases or technologies (54.2\%), and half (50.2\%) hoped agents would make development work feel less mentally demanding.




\subsection{RQ1: How do developers orchestrate their work with AI coding agents?}
\label{sec:rq1}

Developers \textbf{\textit{orchestrated}} agent work through decisions both before and during execution. 
Before execution, they decided whether to delegate, which model to use, and how much autonomy to grant. 
During execution, they monitored the agent and decided when to intervene or ask the agent to check in.
We begin with developers’ delegation decisions.

\noindent\textbf{Delegation decisions.}
Developers first decided whether to delegate at all.
They chose to work without an agent when doing the task themselves would be faster or easier (59\%), when they wanted to understand the problem deeply or practice a skill (51\%), or when further guidance and correction would take more effort than doing the task themselves (38\%).
When they did delegate, model choice was often deliberate: only 27\% usually used the tool's default model, while 73\% matched models to task difficulty, 59\% to specific capabilities, and 41\% used different models at different stages, such as planning and implementation.
Cost and usage limits also shaped use. Only 9\% said these did not affect their use, and developers using free tiers more often used different models by stage than those who did not (57\% vs.\ 31\%).
Prompting effort was most often concentrated in writing or modifying code (45\%), research or learning unfamiliar technologies (43\%), and debugging (42\%).

\finding{1}{Autonomy varies more across developers than across activities.}

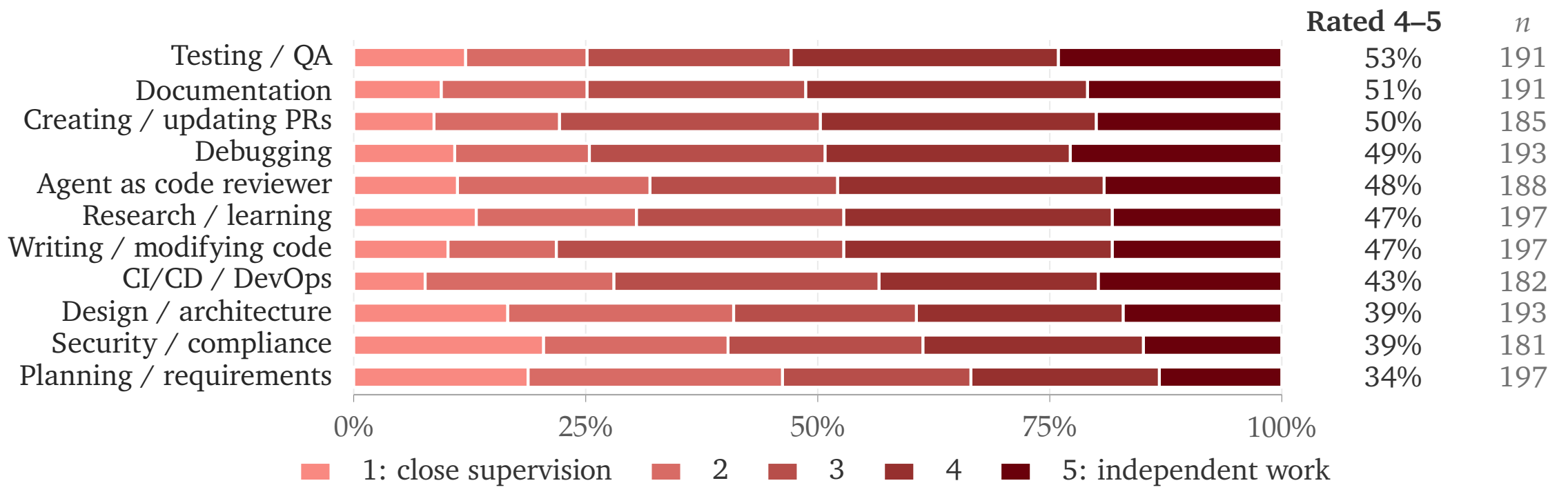
\begin{figure}[tbp]
\centering
\begingroup
\pgfplotstableread{
id c1 c2 c3 c4 c5 n
1 23 25 42 55 46 191
2 18 30 45 58 40 191
3 16 25 52 55 37 185
4 21 28 49 51 44 193
5 21 39 38 54 36 188
6 26 34 44 57 36 197
7 20 23 61 57 36 197
8 14 37 52 43 36 182
9 32 47 38 43 33 193
10 37 36 38 43 27 181
11 37 54 40 40 26 197
}\rqplotdata
\begin{tikzpicture}
\begin{axis}[rq stacked,
height=3.4cm,
ymin=0.45,ymax=11.55,
ytick={1,...,11},
yticklabels={{Testing / QA},{Documentation},{Creating / updating PRs},{Debugging},{Agent as code reviewer},{Research / learning},{Writing / modifying code},{CI/CD / DevOps},{Design / architecture},{Security / compliance},{Planning / requirements}}
]
\addplot[fill=rqlevel1,draw=white,line width=.8pt] table[x expr=100*\thisrow{c1}/\thisrow{n},y=id] {\rqplotdata};
\addlegendentry{1: close supervision}
\addplot[fill=rqlevel2,draw=white,line width=.8pt] table[x expr=100*\thisrow{c2}/\thisrow{n},y=id] {\rqplotdata};
\addlegendentry{2}
\addplot[fill=rqlevel3,draw=white,line width=.8pt] table[x expr=100*\thisrow{c3}/\thisrow{n},y=id] {\rqplotdata};
\addlegendentry{3}
\addplot[fill=rqlevel4,draw=white,line width=.8pt] table[x expr=100*\thisrow{c4}/\thisrow{n},y=id] {\rqplotdata};
\addlegendentry{4}
\addplot[fill=rqlevel5,draw=white,line width=.8pt] table[x expr=100*\thisrow{c5}/\thisrow{n},y=id] {\rqplotdata};
\addlegendentry{5: independent work}
\node[rq head,anchor=south east] at (axis cs:118,0.4) {Rated 4--5};
\node[rq head,text=black!55] at (axis cs:\rqnx,0.4) {$n$};
\node[rq note] at (axis cs:\rqhighx,1) {53\%};
\node[rq note,text=black!55] at (axis cs:\rqnx,1) {191};
\node[rq note] at (axis cs:\rqhighx,2) {51\%};
\node[rq note,text=black!55] at (axis cs:\rqnx,2) {191};
\node[rq note] at (axis cs:\rqhighx,3) {50\%};
\node[rq note,text=black!55] at (axis cs:\rqnx,3) {185};
\node[rq note] at (axis cs:\rqhighx,4) {49\%};
\node[rq note,text=black!55] at (axis cs:\rqnx,4) {193};
\node[rq note] at (axis cs:\rqhighx,5) {48\%};
\node[rq note,text=black!55] at (axis cs:\rqnx,5) {188};
\node[rq note] at (axis cs:\rqhighx,6) {47\%};
\node[rq note,text=black!55] at (axis cs:\rqnx,6) {197};
\node[rq note] at (axis cs:\rqhighx,7) {47\%};
\node[rq note,text=black!55] at (axis cs:\rqnx,7) {197};
\node[rq note] at (axis cs:\rqhighx,8) {43\%};
\node[rq note,text=black!55] at (axis cs:\rqnx,8) {182};
\node[rq note] at (axis cs:\rqhighx,9) {39\%};
\node[rq note,text=black!55] at (axis cs:\rqnx,9) {193};
\node[rq note] at (axis cs:\rqhighx,10) {39\%};
\node[rq note,text=black!55] at (axis cs:\rqnx,10) {181};
\node[rq note] at (axis cs:\rqhighx,11) {34\%};
\node[rq note,text=black!55] at (axis cs:\rqnx,11) {197};
\end{axis}
\end{tikzpicture}
\endgroup
\caption{Reported autonomy across development activities. Darker shades indicate more autonomy; rows are ordered by the percentage rating autonomy 4 or 5, from highest to lowest. $n$ excludes missing and not-applicable responses.}
\label{fig:rq1-autonomy}
\end{figure}

\noindent When developers delegated work to agents, they also decided how much autonomy to grant them~\cite{choudhuri2026you}.
We analyzed autonomy ratings on a five-point scale, from 1 (close supervision) to 5 (independent work)~\cite{feng2025levels}.
Figure~\ref{fig:rq1-autonomy} shows how respondents rated each activity, ordered by the share who gave high autonomy (4 or 5).
Testing (53\%) and documentation (51\%) received high autonomy most often, while planning (34\%) and security and compliance (39\%) received it least often.
Within each activity, however, ratings spanned the full scale: even for testing, 12\% kept close supervision, and even for planning, 13\% let the agent work independently.
This spread within each activity suggests that the developer, not only the activity, shapes how much autonomy agents receive.
To compare these two sources of variation, we fitted a linear mixed-effects model with crossed random intercepts for developers and activities~\cite{baayen2008mixed}, because each developer rated multiple activities and each activity was rated by multiple developers.
We estimated variance components using restricted maximum likelihood and expressed each component as a proportion of the total variance.
We calculated 95\% confidence intervals using 2,000 parametric bootstrap simulations, refitting the model each time.
Across the 11 development activities, reported autonomy varied much more between developers than between activities.
Differences between developers explained 27.6\% of the variance (95\% CI: 22.7--32.7\%), compared with 2.1\% for differences between activities (95\% CI: 0.4--4.8\%).
The remaining 70.3\% was not captured by developers' overall autonomy levels or average differences between activities.
This remainder may reflect how individual developers vary autonomy across activities, other unmeasured factors, and measurement error.
As a sensitivity check, an ordinal mixed model that does not assume equal spacing between ratings~\cite{christensen2019ordinal} gave the same result (29\% vs.\ 2\%).

Thus, although reported autonomy differed across development activities, activity type explained much less variation than differences among developers.
This result complements prior evidence that acceptable levels of AI autonomy can vary across both tasks and individuals~\cite{choudhuri2026you}: in our respondents' reported practices, the stronger source of variation was the developer.

\finding{2}{Developers intervene mainly when agents deviate from task intent, scope, or direction, and their desired check-ins reflect similar boundaries.}

\noindent Developers continued to monitor agents after delegation.
Specifically, 57\% reported that they watched agent output and intervened when needed.
The three most commonly reported reasons for intervening all concerned whether the agent remained aligned with the intended task: the agent misunderstood the task or headed in the wrong direction (65\%), chose a wrong or overcomplicated approach (54\%), or changed more than requested (51\%). Fewer developers intervened because the agent was about to modify files it should not touch (35\%) or run a risky or destructive command (34\%). Similarly, a study of 600 rejected agent pull requests identified agent misalignment and unwanted feature implementations among the reasons for rejection~\cite{ehsani2026ai}.
One respondent described the work needed to correct an agent's assumptions:

\participantquote{P003}{I end up spending more time explaining the codebase and correcting assumptions}

\noindent Beyond intervening while agents worked, developers wanted agents to pause and check in at specific points.
For example, 65\% wanted a check-in before changes to critical, production, or security-sensitive code, and 57\% wanted agents to pause when instructions were ambiguous.
We then compared intervention reasons with check-in preferences for the 196 respondents who answered both questions, testing all 70 reason--check-in pairs (Fisher's exact tests with BH correction; bootstrap 95\% confidence intervals for $\phi$).
Not selecting a reason does not mean it never occurred.
Of the 70 pairs, 14 showed significant positive associations.
We then grouped the significant pairs by intervention reason into \textit{task intent and direction} (3), \textit{scope} (9), and \textit{time/budget} (2).
We highlight one example from each group.
Respondents who selected misunderstanding the task as a reason for intervention more often wanted check-ins when instructions were ambiguous (66\% vs.\ 40\%; $\phi=.244$, 95\% CI [.110, .375], $q=.009$).
Those who selected changes beyond the request more often wanted check-ins when the agent departed from the original plan (59\% vs.\ 34\%; $\phi=.248$, 95\% CI [.117, .388], $q=.008$).
Those who selected excess time or budget more often wanted check-ins when time or token use exceeded expectations (63\% vs.\ 20\%; $\phi=.375$, 95\% CI [.228, .512], $q<.001$).
No pairs involving risky commands or repeated unsuccessful actions were significant.
These results suggest that developers want agents to pause at the points where they have seen agents go wrong.

\rqsummary{1}{
Developers actively orchestrated agent work before and during execution.
Reported autonomy varied much more between developers than between development activities.
After delegation, developers continued to monitor agents and most often intervened when agents deviated from task intent, scope, or direction.
Their desired check-ins corresponded to the problems that prompted intervention.
}
\subsection{RQ2: How do developers assure the work of AI coding agents?}
\label{sec:rq2}

We use \textbf{\textit{assurance}} to describe how developers review and verify agent-generated work and manage security and privacy risks.
We begin with developers' current practices.

\noindent\textbf{Current assurance practices.}
Of the 203 respondents, 63\% reported requiring approval or warnings before risky actions, 53\% restricted agent access, and 50\% required additional human review for high-risk changes. For reviewing agent-generated code, 63\% compared output against requirements, 56\% inspected diffs, and 53\% focused on understanding the approach behind changes. Fewer reported relying on automated tests or continuous integration checks (38\%), another developer (21\%), or another AI tool (12\%).
For pull requests containing substantial AI-generated code, 57\% paid more attention to whether changes were unnecessarily large or complex, while only 7\% reviewed them in roughly the same way as other pull requests.
Developers also reported difficulties when reviewing agent-generated code (n=201). More than half (55\%) found reviewing large or multi-file changes difficult, and 51\% found it difficult to spot problems in plausible-looking code.
Only 2\% reported generally finding agent-generated code review not challenging.
In addition, 69\% (140/202) agreed that line-by-line review could not keep up with agent-generated code.
To our surprise, only 20\% (41/203) reported logging or auditing agent actions.
Prior work similarly found that developers sometimes relied on reviewing final outputs while giving limited attention to action logs during execution~\cite{dhanorkar2026human}.
This motivated us to compare how carefully developers reviewed generated code with how closely they reviewed agent actions before approval.

\finding{3}{Careful code review does not necessarily extend to careful action approval.}

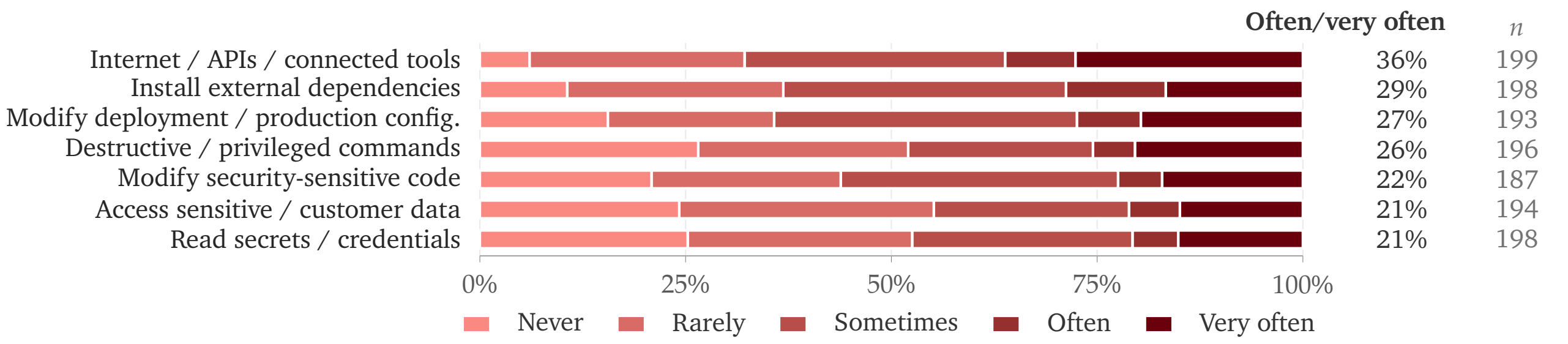
\begin{figure}[tbp]
\centering
\begingroup
\pgfplotstableread{
id c1 c2 c3 c4 c5 n
1 12 52 63 17 55 199
2 21 52 68 24 33 198
3 30 39 71 15 38 193
4 52 50 44 10 40 196
5 39 43 63 10 32 187
6 47 60 46 12 29 194
7 50 54 53 11 30 198
}\rqplotdata
\begin{tikzpicture}
\begin{axis}[rq stacked,
height=2.3cm,
ymin=0.45,ymax=7.55,
ytick={1,...,7},
yticklabels={{Internet / APIs / connected tools},{Install external dependencies},{Modify deployment / production config.},{Destructive / privileged commands},{Modify security-sensitive code},{Access sensitive / customer data},{Read secrets / credentials}}
]
\addplot[fill=rqlevel1,draw=white,line width=.8pt] table[x expr=100*\thisrow{c1}/\thisrow{n},y=id] {\rqplotdata};
\addlegendentry{Never}
\addplot[fill=rqlevel2,draw=white,line width=.8pt] table[x expr=100*\thisrow{c2}/\thisrow{n},y=id] {\rqplotdata};
\addlegendentry{Rarely}
\addplot[fill=rqlevel3,draw=white,line width=.8pt] table[x expr=100*\thisrow{c3}/\thisrow{n},y=id] {\rqplotdata};
\addlegendentry{Sometimes}
\addplot[fill=rqlevel4,draw=white,line width=.8pt] table[x expr=100*\thisrow{c4}/\thisrow{n},y=id] {\rqplotdata};
\addlegendentry{Often}
\addplot[fill=rqlevel5,draw=white,line width=.8pt] table[x expr=100*\thisrow{c5}/\thisrow{n},y=id] {\rqplotdata};
\addlegendentry{Very often}
\node[rq head,anchor=south east] at (axis cs:118,0.4) {Often/very often};
\node[rq head,text=black!55] at (axis cs:\rqnx,0.4) {$n$};
\node[rq note] at (axis cs:\rqhighx,1) {36\%};
\node[rq note,text=black!55] at (axis cs:\rqnx,1) {199};
\node[rq note] at (axis cs:\rqhighx,2) {29\%};
\node[rq note,text=black!55] at (axis cs:\rqnx,2) {198};
\node[rq note] at (axis cs:\rqhighx,3) {27\%};
\node[rq note,text=black!55] at (axis cs:\rqnx,3) {193};
\node[rq note] at (axis cs:\rqhighx,4) {26\%};
\node[rq note,text=black!55] at (axis cs:\rqnx,4) {196};
\node[rq note] at (axis cs:\rqhighx,5) {22\%};
\node[rq note,text=black!55] at (axis cs:\rqnx,5) {187};
\node[rq note] at (axis cs:\rqhighx,6) {21\%};
\node[rq note,text=black!55] at (axis cs:\rqnx,6) {194};
\node[rq note] at (axis cs:\rqhighx,7) {21\%};
\node[rq note,text=black!55] at (axis cs:\rqnx,7) {198};
\end{axis}
\end{tikzpicture}
\endgroup
\caption{Reported frequency of approving agent actions without close review. Darker shades indicate more frequent approval; rows are sorted by the share answering \emph{Often} or \emph{Very often}. $n$ excludes missing and not-applicable responses.}
\label{fig:rq2-action-approval}
\end{figure}

\noindent We asked how often developers approved seven types of agent actions without close review (Figure~\ref{fig:rq2-action-approval}).
Overall, 64\% (130/203) often approved at least one of these action types without close review. The share was lowest for reading secrets or other sensitive files (21\%) and highest for accessing the internet, external APIs, or connected tools (36\%).
Careful code reviewers were not significantly less likely to do so.
Among developers who carefully reviewed most or all agent-generated code (52\%, 105/202), 61\% (64/105) often approved at least one action type without close review, compared with 67\% (65/97) of other developers (OR $=0.77$, 95\% CI: $0.41$--$1.42$; $p=.383$).
Overall code-review habits were also not associated with how often developers approved any of the seven action types without close review after BH correction.

Some individual review practices, however, were related to how often developers approved actions without close review: six of the 76 testable practice--action pairs were significant. 
Developers who inspected diffs approved four action types without close review less often: reading secrets or other sensitive files; accessing personal, customer, proprietary, or regulated data; modifying CI/CD, deployment, infrastructure, database, or production configurations; and running destructive or privileged commands ($r_{\mathrm{rb}}=-.329$ to $-.252$, $q=.004$--$.021$).
Those who checked edge cases approved destructive or privileged commands without close review less often ($r_{\mathrm{rb}}=-.291$, $q=.007$), whereas those who consulted documentation approved reading secrets or other sensitive files without close review more often ($r_{\mathrm{rb}}=.285$, $q=.018$).
Because RQ1 showed that developers differ in how much autonomy they grant, we also asked whether developers who gave agents more autonomy compensated by reviewing actions more closely or by using more safeguards.
We found no sign of such compensation: higher autonomy went with more frequent approval without close review in 48 of the 77 activity--action pairs ($\rho=.16$--$.43$, $q<.05$), and none of the 110 activity--safeguard pairs was significant. Autonomy in planning, architecture, and security and compliance work was associated with approval without close review for all seven action types, whereas autonomy in documentation and research showed no significant associations.
These results suggest that code review and action approval are separate checks: developers who carefully review the code an agent produces do not necessarily look as closely at the actions it asks to perform. Because higher autonomy also went with more frequent approval without close review, approval requests may provide the least scrutiny where developers delegate the most.
One exception was inspecting diffs, which gives developers something concrete to check before they approve.
One respondent described wanting more information before approving:
\participantquote{P198}{I'd also want stronger built-in verification, so the agent could test its work, catch security issues, and clearly explain what it changed before asking me to approve it.}

\finding{4}{Risk awareness and concern do not consistently correspond to more careful approval.}

\noindent Most developers were very or extremely concerned about leaking personal data (68\%) or secrets (68\%), excessive agent permissions (64\%), insecure generated code (64\%), and unsafe commands (62\%).
In addition, 73\% (147/201) were aware of at least one security or privacy incident or concerning practice involving agents, such as agent-generated code containing hard-coded secrets.
This awareness was not associated with more careful approval. Among developers aware of at least one incident or concerning practice, 70\% (103/147) often approved at least one action type without close review, compared with 43\% (20/46) of those who reported no incident or concerns (OR $=3.02$, 95\% CI: 1.46--6.38; Fisher $p=.001$).
Developers who use agents more may both hear of more incidents and approve more actions without close review. 
We fitted a logistic regression~\cite{hosmer2000applied} with incident awareness as the predictor, adjusting for how often and how long developers had used agents and for recruitment channel.
The association remained (adjusted OR $=2.76$, 95\% CI: 1.32--5.78; $n=183$).
The result was similar when we included only developers who rated all seven action types, and the same pattern appeared when we compared each answer to the awareness question with each action type (26 of 56 pairs significant after BH correction).
Developers in safety-critical contexts showed a similar pattern. These 65 developers (32\%) worked on systems where failures could cause physical, medical, or operational harm. They reported more frequent approval without close review  than other developers for six of the seven action types.
Greater concern did go with some privacy practices, however, such as removing sensitive details before prompting (11 of 72 pairs), but not with safeguard use (none of 80 pairs).
These results suggest that awareness of risk does not translate into closer review when agents ask for approval.

\rqsummary{2}{
Developers relied heavily on manual code review, but careful code review did not imply equally careful action approval: among respondents who carefully reviewed most or all generated code, 61\% often approved at least one action type without close review.
Higher reported autonomy was also associated with more frequent approval without close review when approval was requested.
Risk awareness and concern did not consistently correspond to more careful action approval.
}
\subsection{RQ3: How do developers experience working with AI coding agents?}
\label{sec:rq3}

Developers' \textbf{\textit{experience}} with agents includes both the challenges they identified and the improvements they wanted. Despite the benefits they reported, developers described demanding supervision and recurring agent problems. These challenges were associated with how they viewed their work.

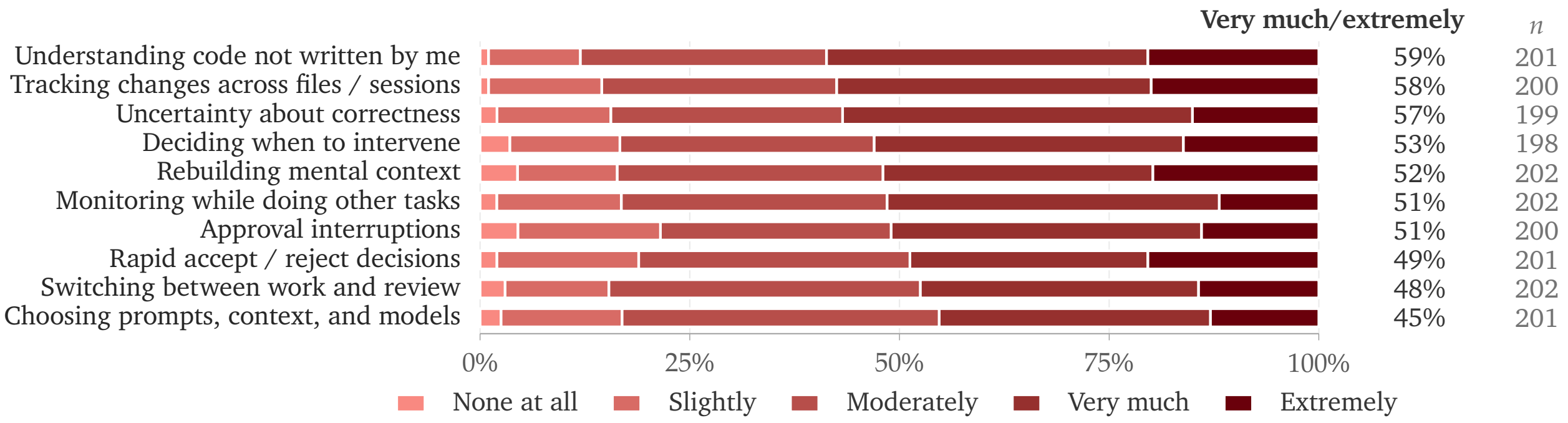
\begin{figure}[tbp]
\centering
\begingroup
\pgfplotstableread{
id c1 c2 c3 c4 c5 n
1 2 22 59 77 41 201
2 2 27 56 75 40 200
3 4 27 55 83 30 199
4 7 26 60 73 32 198
5 9 24 64 65 40 202
6 4 30 64 80 24 202
7 9 34 55 74 28 200
8 4 34 65 57 41 201
9 6 25 75 67 29 202
10 5 29 76 65 26 201
}\rqplotdata
\begin{tikzpicture}
\begin{axis}[rq stacked,
height=3.1cm,
ymin=0.45,ymax=10.55,
ytick={1,...,10},
yticklabels={{Understanding code not written by me},{Tracking changes across files / sessions},{Uncertainty about correctness},{Deciding when to intervene},{Rebuilding mental context},{Monitoring while doing other tasks},{Approval interruptions},{Rapid accept / reject decisions},{Switching between work and review},{Choosing prompts, context, and models}}
]
\addplot[fill=rqlevel1,draw=white,line width=.8pt] table[x expr=100*\thisrow{c1}/\thisrow{n},y=id] {\rqplotdata};
\addlegendentry{None at all}
\addplot[fill=rqlevel2,draw=white,line width=.8pt] table[x expr=100*\thisrow{c2}/\thisrow{n},y=id] {\rqplotdata};
\addlegendentry{Slightly}
\addplot[fill=rqlevel3,draw=white,line width=.8pt] table[x expr=100*\thisrow{c3}/\thisrow{n},y=id] {\rqplotdata};
\addlegendentry{Moderately}
\addplot[fill=rqlevel4,draw=white,line width=.8pt] table[x expr=100*\thisrow{c4}/\thisrow{n},y=id] {\rqplotdata};
\addlegendentry{Very much}
\addplot[fill=rqlevel5,draw=white,line width=.8pt] table[x expr=100*\thisrow{c5}/\thisrow{n},y=id] {\rqplotdata};
\addlegendentry{Extremely}
\node[rq head,anchor=south east] at (axis cs:118,0.4) {Very much/extremely};
\node[rq head,text=black!55] at (axis cs:\rqnx,0.4) {$n$};
\node[rq note] at (axis cs:\rqhighx,1) {59\%};
\node[rq note,text=black!55] at (axis cs:\rqnx,1) {201};
\node[rq note] at (axis cs:\rqhighx,2) {58\%};
\node[rq note,text=black!55] at (axis cs:\rqnx,2) {200};
\node[rq note] at (axis cs:\rqhighx,3) {57\%};
\node[rq note,text=black!55] at (axis cs:\rqnx,3) {199};
\node[rq note] at (axis cs:\rqhighx,4) {53\%};
\node[rq note,text=black!55] at (axis cs:\rqnx,4) {198};
\node[rq note] at (axis cs:\rqhighx,5) {52\%};
\node[rq note,text=black!55] at (axis cs:\rqnx,5) {202};
\node[rq note] at (axis cs:\rqhighx,6) {51\%};
\node[rq note,text=black!55] at (axis cs:\rqnx,6) {202};
\node[rq note] at (axis cs:\rqhighx,7) {51\%};
\node[rq note,text=black!55] at (axis cs:\rqnx,7) {200};
\node[rq note] at (axis cs:\rqhighx,8) {49\%};
\node[rq note,text=black!55] at (axis cs:\rqnx,8) {201};
\node[rq note] at (axis cs:\rqhighx,9) {48\%};
\node[rq note,text=black!55] at (axis cs:\rqnx,9) {202};
\node[rq note] at (axis cs:\rqhighx,10) {45\%};
\node[rq note,text=black!55] at (axis cs:\rqnx,10) {201};
\end{axis}
\end{tikzpicture}
\endgroup
\caption{Reported mental demand of supervision activities. Darker shades indicate greater demand; rows are sorted by the share answering \emph{Very much} or \emph{Extremely}. $n$ excludes missing and not-applicable responses.}
\label{fig:rq3-supervision-demand}
\end{figure}

\noindent\textbf{Workload savings alongside demanding supervision.}
Most developers reported benefits: agents reduced their workload (83\%) and the mental effort required for tasks (72\%), justified their monetary cost (76\%), and made coding more enjoyable (72\%).
These benefits, however, did not imply that supervising agents was easy.
Overall, 63\% (128/203) rated at least five of 10 supervision activities as very or extremely mentally demanding.
Half of all respondents (101/203) both agreed that agents reduced their workload and rated at least five supervision activities as very or extremely demanding.
The activities most often rated as very or extremely demanding were understanding and reviewing code they had not written (59\%), tracking changes across files, sessions, or agents (58\%), and dealing with uncertainty about output correctness and completeness (57\%) (Figure~\ref{fig:rq3-supervision-demand}).
All three involve understanding and checking work the developer did not produce.
Understanding changes is a known challenge in code review~\cite{bacchelli2013expectations}, and verifying AI suggestions takes a large share of programmers' time~\cite{mozannar2024reading}; with agents, this checking spans many files and sessions.

\noindent\textbf{Agent problems and workflow disruptions.}
Developers also reported recurring agent problems.
They often encountered agents claiming completion while parts were missing or broken (41\%), over-engineering solutions (41\%), or producing plausible but incorrect output (39\%).
These problems mean that agent output cannot be taken at face value, which adds to the supervision demands above.
Over time, 65\% noticed code becoming more complex, bloated, or hard to maintain.
Workflow disruptions were also frequent: 37\% often restarted sessions that became too long or confused, 39\% hit usage limits before completing a task, and 25\% lost progress after unexpected context resets.

\finding{5}{Supervision demands and agent problems are associated with strain, while some demands are also associated with approving code that is not fully understood.}

\noindent Supervision demands and agent problems are both costs of working with agents, and prior work suggests such costs can affect satisfaction and well-being~\cite{forsgren2021space,wu2026developers}.
We therefore correlated how demanding developers found each of 10 supervision activities, and how often they encountered each of 12 agent problems, with how much they agreed with nine statements about their views of their work, such as ``AI coding agents have reduced my satisfaction with the work'' and ``I am concerned that AI coding agents may reduce demand for developers in roles like mine.''
Of these 198 pairs, 90 were significant after BH correction.
We group the results by statement.

Most significant pairs involved four statements, on reduced satisfaction, self-doubt, job concern, and pressure to keep up with others (73 of the 88 pairs involving these statements), for both supervision demands and agent problems.
For example, developers who found monitoring agents while doing other tasks more demanding tended to agree more that agents had reduced their satisfaction with the work ($\rho=.337$, 95\% CI [.201, .466], $q=.002$) and that others' productivity with agents made them doubt their own way of working ($\rho=.371$, 95\% CI [.230, .499], $q=.002$).
One respondent described this change:
\participantquote{P047}{the psychological shift from actively writing code to mostly reviewing and editing code, which changes how engineering fatigue feels over a long work week}

\noindent Whether developers felt agents were worth the money depended mainly on agent problems.
For 11 of the 12 agent problems, developers who encountered the problem more often agreed less that agents were worth their monetary cost; this was true for only one of the 10 supervision demands.
For example, developers who more often received plausible but incorrect output tended to agree less that agents were worth their monetary cost ($\rho=-.261$, 95\% CI [$-$.387, $-$.122], $q=.007$).
\participantquote{P048}{the AI is adding more work than it solves since it adds so many lines of code per change. Unfortunately we are very pressured to use it.}

\noindent Overall, supervision demands and agent problems went with more concern and less satisfaction, but much less with enjoyment or ownership (5 of 88 pairs): these costs appear to sit alongside what developers value about working with agents rather than displace it. For example, about one-quarter of respondents (49/198) agreed both that agents made coding more enjoyable and that agents reduced their satisfaction with the work.
This pattern fits the Job Demands-Resources model, in which job demands mainly lead to strain, whereas engagement depends mainly on job resources~\cite{bakker2007job}.
This coexistence is not new to agents: in an earlier Copilot study, developers preferred the tool even though its code was hard to understand and debug~\cite{vaithilingam2022expectation}.

Supervision demands may also affect what developers accept.
Half of respondents (50\%) agreed that they sometimes approved agent-generated code they did not fully understand because rewriting it would take too long, and 46\% worried about being held responsible for code they did not fully understand or no longer felt was theirs.
Developers who found two of the 10 supervision activities more demanding agreed more that they approved such code: monitoring agents while doing other tasks ($\rho=.236$, 95\% CI [.098, .375], $q=.040$) and handling approval requests and other interruptions ($\rho=.224$, 95\% CI [.084, .359], $q=.043$).
Developers who agreed more that they approved such code also felt more pressure to keep up with others ($\rho=.207$, 95\% CI [.063, .347], $q=.007$).
This fits findings that people monitor automation less closely when busy with other tasks~\cite{parasuraman2010complacency}.

These challenges raise the question of what would make developers willing to give agents more autonomy. We asked this in an open-ended question (175 coded responses) and also asked what one thing developers would fix or add to their agents (97 coded responses).

\finding{6}{Beyond reliability, developers want autonomy that is bounded, observable, and reversible.}

\noindent One respondent described the autonomy they would accept in three words:
\participantquote{P017}{making autonomy bounded, observable and reversible. I'd be much comfortable giving a coding agent broad freedom if it were difficult for the agent to create a large, silent failure.}
\noindent We adopted these words to name three of the four categories we derived from the responses about granting more autonomy (Table~\ref{tab:rq4-assurance}).
Besides wanting agents to be \emph{reliable} (115/175, 66\%), developers wanted them to be \emph{observable}, showing what they do and why (61/175, 35\%).
They also wanted agents to be \emph{bounded}, working within permissions, security safeguards, and limits on sensitive data (60/175, 34\%), and \emph{reversible}, so that the damage of mistakes could be limited or undone (34/175, 19\%).
In total, 104 responses (59\%) raised at least one of these three categories.

\begin{table}[tb]
\caption{Categories in responses about granting more autonomy (S2.5, $n=175$) and about the one thing developers would fix or add (S7.12, $n=97$).}
\label{tab:rq4-assurance}
\centering
\tiny
\setlength{\tabcolsep}{3pt}
\renewcommand{\arraystretch}{1.15}
\begin{tabularx}{\linewidth}{@{}>{\raggedright\arraybackslash}p{.10\linewidth}>{\raggedright\arraybackslash}p{.30\linewidth}>{\raggedright\arraybackslash}p{.08\linewidth}>{\raggedright\arraybackslash}p{.08\linewidth}>{\raggedright\arraybackslash}X@{}}
\toprule
Category & Main concepts & Autonomy & One thing & Illustrative excerpt (autonomy question) \\
\midrule
Reliable & Getting the work right; checking its own work; fitting changes to the project; remembering project context & 115 (66\%) & 58 (60\%) & \textit{``It understands my codebase, not just code''} (P131) \\
\addlinespace[2pt]
Observable & Showing what it does and why; asking instead of guessing; communicating candidly & 61 (35\%) & 13 (13\%) & \textit{``a persistent, reviewable action log''} (P009) \\
\addlinespace[2pt]
Bounded & Adjustable human supervision, such as permissions and approval checkpoints; security safeguards; protecting sensitive data & 60 (34\%) & 11 (11\%) & \textit{``I want to decide exactly what the agent can access, modify, delete, or deploy without having to supervise every action.''} (P149) \\
\addlinespace[2pt]
Reversible & Limiting the damage of mistakes, such as rollback, checkpoints, and sandboxes & 34 (19\%) & 8 (8\%) & \textit{``every change could be cleanly reverted when something goes wrong''} (P147) \\
\bottomrule
\end{tabularx}
\par\smallskip
\begin{minipage}{\linewidth}\footnotesize
A response could fall into several categories. In the autonomy question, 104 responses (59\%) raised at least one of bounded, observable, or reversible. Percentages describe coded responses, not all survey participants. Quotes are verbatim excerpts.
\end{minipage}
\end{table}

\noindent The question wording also mattered.
Among the 90 developers who answered both questions, bounded (32\% vs.\ 12\%) and observable (28\% vs.\ 14\%) came up more often when developers were asked about autonomy, whereas reliable (64\% vs.\ 61\%) and reversible (13\% vs.\ 9\%) came up about equally often (exact McNemar tests~\cite{mcnemar1947note} with BH correction).
Part of this difference may reflect that the one-thing question asked for a single change.

\rqsummary{3}{
Developers reported benefits from coding agents, such as reduced workload, but these benefits often came with demanding supervision and recurring agent problems.
Supervision demands and agent problems were associated with reduced satisfaction, self-doubt, job concern, and pressure to keep up, and developers who found monitoring and interruptions more demanding agreed more that they approved code they did not fully understand.
Beyond reliable task completion, 59\% of responses about granting more autonomy asked for agents that are bounded, observable, or reversible.
}

\subsection{Robustness Across Recruitment Channels}
\label{sec:channel-consistency}

Data quality and participant backgrounds can differ between recruitment
platforms~\cite{peer2022data,danilova2021you}. As a post-hoc check,
we repeated the quantitative analyses behind Findings 1--5 separately for the community channel (124 respondents) and Prolific (79 respondents).
The key estimates for every finding pointed in the same direction in both channels, and none differed significantly between them. Findings 1 and 3, and the associations with views of work in Finding 5, held in both
channels. For Findings 2 and 4, and for the association with approving
code that is not fully understood in Finding 5, some community-channel
estimates were too uncertain to rule out no effect, likely because of
smaller groups or weaker associations. We did not require significance
in both channels, because each is smaller than the full
sample~\cite{gelman2006difference}. 

\section{Discussion}
\subsection{Three Tensions in Developer-Agent Collaboration}
\label{sec:discussion-autonomy}

Our findings (Section~\ref{sec:results}) reveal three tensions in developer-agent collaboration.
Each tension relates to one of three needs in human-AI collaboration~\cite{klien2004ten}: directing the agent (orchestration), predicting what it will do (assurance), and keeping the effort of working with it manageable (experience). Together, these tensions show that giving agents more autonomy does not take developers out of the work; it moves their effort from writing code to deciding what to delegate and what to approve.

\noindent$\bullet$ \textbf{\textit{Tension \#1: Orchestration: Autonomy Follows the Developer, Corrections Follow Intent.}}
Directing an agent starts with deciding how much autonomy to give it.
Automation research suggests matching autonomy to the consequences of errors~\cite{parasuraman2000model}, and developers in prior work said they would accept less autonomy when they were responsible for the outcome~\cite{choudhuri2026you}.
Our results only partly support this view: autonomy differed far more between developers than between activities (Finding~1).
One reason may be that accepting autonomy in principle differs from granting it in practice, where habit and convenience may matter more than the stakes of each task.
Agent design may reinforce this pattern, since many agents set permissions for a whole session or project rather than for each kind of work.
A second pattern appears in why developers intervene.
They mostly stepped in because the agent misunderstood the task, chose a wrong or overcomplicated approach, or changed more than requested, and fewer did so because it was about to run a risky command (Finding~2).
An autonomy level determines how independently an agent can work, but not whether it has understood what the developer wants.
Klein et al.\ argue that joint work depends on a shared understanding of the goal that both parties maintain and repair~\cite{klien2004ten}. Consistent with this idea, developers who had corrected misunderstood tasks more often wanted agents to check in when instructions were ambiguous (Finding~2).
Directing an agent may therefore depend less on choosing an autonomy level than on building and maintaining a shared understanding of the goal.

\noindent$\bullet$ \textbf{\textit{Tension \#2: Assurance: Checking What Agents Write, Not What They Do.}}
Action approval requires developers to assess what an agent is about to do.
Code review is an after-the-fact check of an artifact that appears as a diff and can be revisited~\cite{sadowski2018modern}, whereas action approval asks developers to predict, often mid-task, the effects of something that has not yet happened and may never appear in the code.
Huang et al.\ found that experienced developers control agents through planning and close supervision~\cite{huang2025professional}; our results suggest that this control may stop at the code (Finding~3).
Requiring approval before risky actions was the safeguard developers used most (63\%).
Yet 64\% often approved at least one action type without close review.
Developers who inspected diffs reviewed actions more closely, perhaps because a diff gives them something concrete to check.
Developers aware of security or privacy incidents also approved without close review more often, even after we controlled for how often and how long they had used agents (Finding~4).
One possibility is that developers shift attention from approval to monitoring: 57\% watched agent output and intervened when needed (Finding~2), and experienced Claude Code users have been reported to both allow automatic approval and interrupt agents more often~\cite{anthropic2026autonomy}. Repeated approval requests may also receive less attention over time~\cite{anderson2015polymorphic,felt2012android}.
Future studies that measure developers' attention and the number of approval requests could test these explanations.
In short, the care developers give to code does not necessarily extend to the actions agents take, even among developers who know the risks.

\noindent$\bullet$ \textbf{\textit{Tension \#3: Experience: Lighter Workload, Demanding Supervision.}}
All of this takes effort.
Most developers reported a lighter workload (83\%), yet 63\% found at least half of the supervision activities very or extremely demanding, especially understanding and checking work they had not written (RQ3).
As Bainbridge observed, automation can leave people with the hardest part of the work: monitoring a system they no longer operate~\cite{bainbridge1983ironies}.
With agents, this checking spans files and sessions, and agents' own reports are not always reliable: 41\% often saw agents claim completion while parts were missing or broken.
These demands also went with self-doubt, job concern, and, for some demands, approving code developers did not fully understand (Finding~5).
When asked what would make them give agents more autonomy, rather than what they would improve in general, developers more often wanted agents to be bounded and observable (Finding~6).
Greater autonomy may therefore depend not only on agent reliability, but also on being able to see, limit, and undo what agents do.
More reliable agents alone may not be enough.
Bainbridge noted that the less often a system fails, the less practiced people become at catching the failures that remain~\cite{bainbridge1983ironies}.
As agents become more reliable, being able to see and undo what agents do may therefore matter more. 

\subsection{Implications}
\label{sec:discussion-implications}

\noindent\textbf{For Agent Builders.}
Our findings suggest four directions for agent design, following the order of our research questions.
Future studies could assess whether these directions improve approval decisions and reduce supervision effort:

\begin{enumerate}[label=\textbf{D\arabic*.}, wide=0pt]
    \item \textbf{Confirm intent before acting:}
Agents could confirm the goal and scope before starting a task, rather than only asking for approval of individual actions during execution.
    \item \textbf{Match permissions to action risk:}
Instead of applying one rule to everything in a session, agents could apply different rules to different types of actions: low-risk actions proceed automatically, risky actions require approval, and some actions are never allowed.
    \item \textbf{Show what an action will change:}
Agents could preview which files or resources an action will touch and what it will change, then keep a record that developers can review as easily as a code diff.
    \item \textbf{Make mistakes recoverable:}
Agents could provide checkpoints and rollback for both code changes and other actions, allowing developers to undo unwanted effects instead of catching every problem before approval.
\end{enumerate}

\noindent\textbf{For Practitioners.}
Our findings also suggest three practices for teams that work with agents:

\begin{enumerate}[label=\textbf{P\arabic*.}, wide=0pt]
    \item \textbf{Agree on agent permissions as a team:}
Because autonomy varied much more between developers than between activities, teams should not leave agent permissions entirely to individual habits.
They can agree on which actions may proceed freely, which require approval, and who checks their effects.
    \item \textbf{Review agent actions, not only code:}
Teams should treat code review and action approval as separate checks.
Only 20\% of developers logged or audited agent actions (Section~\ref{sec:rq2}), so teams could keep such records and review them alongside the resulting code changes. 
    \item \textbf{Count supervision as part of the work:}
Most developers found supervising agents demanding even when their workload decreased (Section~\ref{sec:rq3}), so managers should count supervision as work and leave enough time for review.
\end{enumerate}

\noindent\textbf{For Researchers.}
Our results are exploratory, so we state the main open questions as hypotheses for future studies:

\begin{enumerate}[label=\textbf{H\arabic*.}, wide=0pt]
    \item Approval requests that preview an action's effects lead to closer review and better decisions, not only to more confidence.
    \item Developers who watch agent output and interrupt when needed review approval requests less closely.
    \item Developers review approval requests less closely as the number of requests in a session grows.
    \item Demanding supervision leads developers to approve code they do not fully understand, rather than the reverse.

\end{enumerate}
Studies of productivity with agents should also count supervision effort, not only output~\cite{forsgren2021space}.

\section{Threats to Validity}
\noindent\textbf{Construct validity.}
Our results rely on self-reports, which may be affected by recall errors, social desirability, and differences in how respondents understood the questions. To reduce \textit{recall errors}, we asked respondents to think about the past month or their most recent typical work month. To reduce \textit{social desirability bias}, the survey was anonymous: we did not ask for names, and community-channel respondents who wanted compensation gave an email address that we stored separately from their answers. To reduce \textit{misunderstanding}, we ran think-aloud sessions with three participants, piloted the survey with another 15, and revised unclear questions and response options (Section~\ref{sec:survey-design}). We also defined key terms in the questions, for example by describing what 1 and 5 meant on the autonomy scale, which replaced an earlier five-level framework~\cite{feng2025levels} that pilot participants found hard to interpret.
Beyond these general issues, the question on action approval, which Findings~3 and~4 rely on, has further limits. Agents differ in which actions require approval, and developers can pre-approve some actions through settings such as allowlists. We asked whether developers restricted agent access or required approval before risky actions, which partly captures these settings, but we did not ask about allowlists. Approving an action without close review can also be reasonable for familiar or low-risk actions, so we reported each action type separately and treated approval without close review as a reported habit, not as a mistake. Finally, autonomy and approval without close review may overlap in meaning. However, we asked about autonomy for development activities and about approval for action types, and autonomy in documentation and research was not correlated with approval, which suggests that the two questions did not measure the same construct.

\noindent\textbf{Internal validity.}
Our survey collected responses at one point in time, so the associations we report do not establish cause and effect; for example, we cannot tell whether risk awareness came before approval habits. Factors we did not measure, such as workload or task type, may also affect both sides of an association, although the association in Finding~4 remained after we controlled for how often and how long developers had used agents.
A general tendency to agree with statements or select many options could also make unrelated answers appear correlated. Several patterns make this less likely: most significant associations in Finding~2 involved questions with related content, autonomy in documentation and research was not correlated with approval (Finding~3), and some associations in Finding~5 were negative. In Finding~4, excluding the 12 respondents who gave the same answer for all seven action types did not change the association.
Careless or fraudulent responses could also distort our results~\cite{danilova2021you}, especially now that bots and AI agents can complete online surveys in ways that are hard to tell apart from human answers~\cite{westwood2025potential,zhang2022beyond}. We therefore screened responses strictly (Section~\ref{sec:screening}), which removed about two-thirds of the community-channel responses that passed the rule-based checks. This screening may have removed some valid responses and kept some fraudulent ones, and detecting AI-generated responses remains an open challenge for future survey research~\cite{westwood2025potential,veselovsky2023artificial}. The time threshold removed no Prolific responses, and the key estimates pointed in the same direction in both channels (Section~\ref{sec:channel-consistency}), which reduces, but does not remove, the risk that the threshold distorted our findings.

\noindent\textbf{Conclusion validity.}
Testing many relationships increases the chance of false positives. To reduce this threat, we applied Benjamini--Hochberg corrections within each analysis set and reported effect sizes with confidence intervals alongside the tests. Some groups were small, so a nonsignificant result does not show that no relationship exists. When a result was significant in one group but not another, we compared the direction and size of the estimates rather than significance alone. To check the stability of key results, we refit Finding~1 with an ordinal model, repeated Finding~4 with controls for agent use and without respondents who gave the same answer for all action types, and compared key estimates across the two recruitment channels (Section~\ref{sec:channel-consistency}). These checks supported our results, but the associations remain exploratory and need confirmation in other samples.

\noindent\textbf{External validity.}
We studied current coding-agent users, so our results do not describe developers who have stopped using agents or never used them, and developers who chose to take part may be more interested in agents than others. Respondents came from 25 countries, although most (64\%) lived in the United States, and Claude Code was the most used agent. Because our study is exploratory, we do not claim that the results generalize to all developers or agents. Recruiting through two channels also reduced our reliance on one source (Section~\ref{sec:channel-consistency}). Finally, we collected responses in August and September 2026 and asked about the past month, so our results reflect practices at that time, and these practices may change as agents evolve. Our findings are about how developers orchestrate, assure, and experience their work with agents, rather than about what a particular agent can do, so we expect them to remain relevant as agents improve.

\section{Conclusion and Future Work}

We surveyed 203 developers about how they orchestrate, assure, and experience their work with AI coding agents, and found three tensions.
First, how much autonomy developers gave agents depended far more on the developer than on the development activity, and when developers stepped in, they mostly corrected misunderstood tasks rather than stopping risky actions.
Second, careful code review did not necessarily extend to action approval: 61\% of developers who carefully reviewed most or all generated code still often approved at least one type of agent action without close review.
Third, agents reduced developers' workload, but supervising them remained mentally demanding.
When asked what would make them give agents more autonomy, most developers wanted agents to complete work reliably, and more than half also wanted agents that are bounded, observable, or reversible.
These tensions suggest that as agents take on more work, developers' effort moves from writing code to deciding what to delegate and what to approve.
Our findings point to four directions for agent design: confirming intent before acting, setting defaults by action risk, showing what an action will change, and making mistakes recoverable.
Future work could investigate whether these designs improve approval decisions and reduce supervision effort, and why developers approve actions without close review.

\bibliography{main}

\end{document}